\documentclass[12pt]{iopart}
\usepackage{iopams}
\usepackage{graphicx}
\def\bB{\mathbf{B}}

\def\bH{\mathbf{H}}

\def\bM{\mathbf{M}}

\def\bi{\mathbf{i}}

\def\bk{\mathbf{k}}

\def\br{\mathbf{r}}

\def\sc{\mathsf{c}}

\def\sf{\mathsf{f}}

\def\sl{\mathsf{l}}

\def\Xint#1{\mathchoice
   {\XXint\displaystyle\textstyle{#1}}%
   {\XXint\textstyle\scriptstyle{#1}}%
   {\XXint\scriptstyle\scriptscriptstyle{#1}}%
   {\XXint\scriptscriptstyle\scriptscriptstyle{#1}}%
   \!\int}
\def\XXint#1#2#3{{\setbox0=\hbox{$#1{#2#3}{\int}$}
     \vcenter{\hbox{$#2#3$}}\kern-.5\wd0}}

\begin{document}

\title{Towards Quantitative Magnetisation Mapping}

\author{Rob F. Remis$^1$ and Peter M. van den Berg$^2$}
\address{$^1$ Circuits and Systems Group, Microelectronics Department, Delft University of Technology, Van Mourik Broekmanweg~6, 2628 XE Delft, The Netherlands}
\address{$^2$ Professor Emeritus, Faculty of Applied Sciences, Delft University of Technology, The Netherlands}
\ead{r.f.remis@tudelft.nl, p.m.vandenberg@tudelft.nl}
\vspace{10pt}
\begin{indented}
\item[] 5 September 2021
\end{indented}

\begin{abstract}
The starting point in quantitative susceptibility mapping (QSM) is a theoretical model that is used to map susceptibility distributions from magnetic field measurements. It requires regularisation techniques to avoid artefacts in the resulting image.
The underlying problem is that the model was developed by starting with the so-called Lorentz sphere on a microscopic scale. After averaging over a macroscopic sample, it is assumed that the magnetic flux density vanishes in the center of the sample. For the macroscopic problem of a homogeneous sphere in a uniform field, we show that at the surface the normal component of the flux density is not continuous, which contradicts Maxwell's macroscopic theory.
In this paper, we propose a model consistent with macroscopic magnetic field theory, in which we image magnetisation rather than susceptibility. This model is well-posed. Some simple but representative numerical examples show that it allows for high-resolution images.
\end{abstract}

\vspace{2pc}
\noindent{\it Keywords}: Quantitative Susceptibility Mapping, Susceptibility, magnetisation, Imaging
%\maketitle

\section{Introduction}

In Quantitative Susceptibility Mapping or QSM, the objective is to image the susceptibility profile within a part of the human body~\cite{Chung&Ruthotto},  \cite{Deistung}, \cite{Ruetten_etal}, \cite{Schweser}. Its development goes back more than 20 years ago~\cite{Schenck} and QSM finds many applications in neuroradiology and neuroimaging, such as
MR venography, oxygen saturation imaging, traumatic brain injury imaging, multiple sclerosis, and brain tumor imaging~\cite{Reichenbach_etal}.

As is well known, the magnetic susceptibility is a constitutive parameter that relates the magnetisation $\bM$ to the magnetic field $\bH$ (to avoid confusion, we refer to $\bH$ as the magnetic field and to $\bB$ as the magnetic flux density). The spatial domain data model that is used in QSM to retrieve this constitutive parameter is obtained by starting from the classic field expression of a magnetic dipole \cite{Jackson} and taking the microscopic Lorentz correction into account~\cite{LinLeigh}. The model can also be formulated in the spatial spectral domain (see \cite{Wang&Liu}, for example). In either case, the QSM problem of retrieving the susceptibility from magnetic field data is ill-posed and requires regularisation \cite{LiuSpincemaille}.  A mathematical analysis of the QSM model, the effects of particular regularisation strategies, and a discussion on the causes of streaking artefacts in reconstructed susceptibility maps  are given in \cite{Choi}.

In the  present paper, we focus on the fundamental QSM model and show that within this model the boundary condition for the magnetic flux density is violated. Specifically, we show that the normal component of the magnetic flux density actually jumps across a source-free interface, where the magnetisation exhibits a jump. Clearly, this is not in accordance with Maxwell's field theory.

Furthermore, in contrast to QSM, we propose to image the magnetisation of tissue rather than its susceptibility. We call this Quantitative Magnetisation Mapping (QMM) and we consider this mapping problem in the spatial domain. Similar to QSM, we assume that only scalar magnetic field data in the direction of the background field (usually the $z$-direction) is available. We show that in the QMM model the boundary conditions for the magnetic field $\bH$ and magnetic flux density $\bB$ are not violated. Moreover, we show that the QMM data operator is actually a shifted version of the QSM data operator and that the ill-posedness of the QSM model does not carry over to the QMM model.

\begin{figure}[t]\hspace*{2cm}
\includegraphics[scale=0.65, viewport = 0 280 700 690,clip =true]  {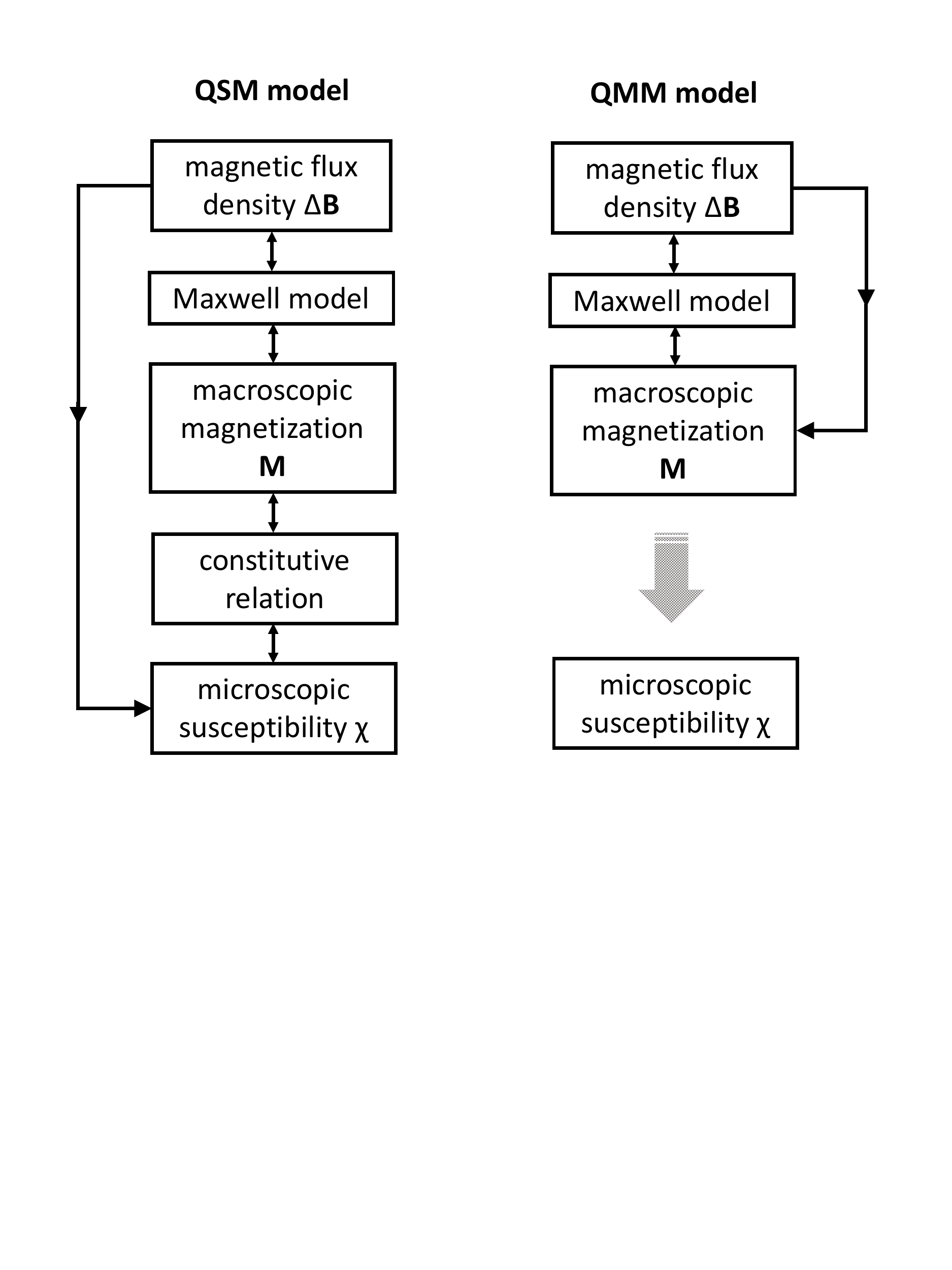}\\[-2mm]
\caption{Schematic overview of the QSM procedure (left  column) and the QMM procedure (right column).}
\label{Figuremap}
\end{figure}

Having a tissue magnetisation map available from QMM, the corresponding magnetic field can be determined. In case a model for the susceptibility function is available from microscopic or quantum mechanical considerations, the susceptibility function can be determined, because the magnetisation and the magnetic field are known. The QMM model does not depend on the particular microscopic or quantum mechanical susceptibility model that it used, since it reconstructs the macroscopic magnetisation from which the corresponding magnetic field can be computed. The susceptibility model is only required to relate the magnetisation to the magnetic field. Figure \ref{Figuremap} gives a schematic overview of QSM and QMM, illustrating the difference between the two approaches.

This paper is organised as follows. In section~\ref{sec:QSM}, we briefly consider the basic QSM model in the spatial domain and show a number of reconstruction results based on simulated and analytical data. These results verify that the QSM operator is indeed ill-posed and regularisation is required to obtain a unique and stable approximate solution. Furthermore, we show that within the QSM model, the boundary condition for the normal component of the magnetic flux density is not satisfied. Subsequently, in  section~\ref{sec:QMM}, we present our QMM approach and demonstrate that within the QMM model the boundary conditions for the magnetic field $\bH$ and magnetic flux density $\bB$ are satisfied at an interface between two media with a different magnetisation. In section~\ref{sec:const}, numerical experiments demonstrating the performance of QMM are presented as well. In addition, we discuss how a susceptibility function of a particular material or tissue type can be reconstructed in QMM. Finally, in section~\ref{sec:compare}, we compare the QSM and QMM operators and show how they are connected. The conclusions can be found in section~\ref{sec:concl}.

\section{Quantitative susceptibility mapping without regularisation}
 \label{sec:QSM}

For each location with position vector $\br = (x,y,z)$, the mathematical problem in QSM is to solve
the susceptibility $\chi(\br)$ from the integral equation \cite{Wang&Liu}
\begin{equation}
\label{eq:QSMspace1}
\Delta B(\br) = {\Xint-\!\!}_{\br'\in \mathbb{D}}\; d(\br\!-\!\br') \,\chi(\br')\, {\rm d}V.
\end{equation}
Here, $\mathbb{D}\in \mathbb{R}^3$ is the domain where $\chi(\br)\neq 0$. Further, the integral ${\Xint-}$ is a Cauchy principal value integral and $\Delta B$ is measured magnetic flux density data due to the presence of  a nonzero susceptibility $\chi$. Finally, $d$ is the so-called dipole kernel, which is given by
\begin{equation}
\label{eq:QSMspace2}
d(\br\!-\!\br') = \frac{3(z\!-\!z')^2 - |\br\!-\!\br'|^2}{|\br\!-\!\br'|^5}
= \frac{\partial^2}{\partial z^2} \frac{1}{|\br\!-\!\br'|},
\end{equation}
for $\br \neq \br'$. If we substitute the right-hand side of (\ref{eq:QSMspace2}) into (\ref{eq:QSMspace1}) and interchange the order of integration and differentiation, we obtain
\begin{equation}
\label{eq:FirstKindequation}
\Delta B(\br) =
\frac{\partial^2}{\partial z^2} {\Xint-\!\!}_{\br'\in \mathbb{D}}\; \frac{1}{|\br\!-\!\br'|} \,\chi(\br')\, {\rm d}V.
\end{equation}
Including the integration point $\br=\br'$ then leads to the integral equation
\begin{equation}
\label{eq:QSMspacemod}
\Delta B(\br) = \frac{1}{3} \chi(\br) +\frac{\partial^2}{\partial z^2}\! \int_{\br'\in \mathbb{D}}\; \frac{1}{|\br\!-\!\br'|} \,\chi(\br')\, {\rm d}V,
 \end{equation}
where we have used that the principal value of the integral is equal to $-\frac{1}{3} \chi(\br)$.
Note that in the spectral domain with spectral vector $\bk = (k_x,k_y,k_z)$, this equation becomes algebraic, cf. \cite{Wang&Liu},
  \begin{equation}
 \label{eq:QSMspaceKspace}
\widetilde{\Delta B}(\bk) = \left[\frac{1}{3} -\frac{k_z^2}{\bk\cdot\bk} \right] \widetilde{\chi}(\bk),
 \end{equation}
 where $\widetilde{\Delta B}$ and $\widetilde{\chi}$ are the spectral domain counterparts of $\Delta B$ and $\chi$, which is clearly not invertible for all $\bk \in \mathbb{R}^3$ satisfying $k_z^2 = \frac{1}{3}\bk\cdot\bk$.

In the following section, our goal is to check the inversion of (\ref{eq:QSMspacemod}) in the spatial domain. Specifically, we first solve a forward problem by computing data $\Delta B$ from QSM equation (\ref{eq:QSMspacemod}) for a known susceptibility~$\chi$. Subsequently, we take this data as input and try to retrieve the susceptibility again using the same QSM equation. Although we then commit an {\em ``inverse crime''} (see p.~154 of Colton and Kress \cite{Colton}), this does allow us to check how well the inverted susceptibility reproduces the input of the forward problem. Finally, we also consider the case where we have analytic data for a homogeneous sphere and use this data to reconstruct the susceptibility of the sphere again using the QSM equation (\ref{eq:QSMspacemod}).

\subsection{Iterative inversion based on circular convolutions}

Equation (\ref{eq:QSMspacemod}) is a convolution type integral equation for the unknown magnetisation. Zwamborn and Van den Berg \cite{Zwamborn} have shown that such an integral equation can efficiently be solved using a weak form formulation together with the so-called iterative CGFFT method. After a spatial discretisation procedure, the discrete spatial convolution is computed as a circular convolution on an extended grid with zero padding. Although sampling in each Cartesian direction is then doubled, the discrete convolution is computed exactly. Here we follow a similar approach, but instead of using the conjugate gradient method, we use BiCGSTAB \cite{VanderVorst} as an iterative solver, since it has been shown that this method typically exhibits fast convergence for well-conditioned convolution type operators \cite{VandenBergBook}. We realise, of course, that the QSM system is severely ill-conditioned, but for a fair comparison of QSM and QMM, we still use the BiCGSTAB method in QSM.
In the discretized formulation, we use a regular grid of $128\times128\times128$ nonoverlapping subdomains (voxels) with a sampling width of 0.002~m. On each subdomain we assume that the susceptibility $\chi$ is constant, and we replace the singular Green function $1/{4\pi|\br\!-\!\br'|}$ by its spherical mean (weak form) \cite{VandenBergBook}. In addition, we keep the ${\partial^2}/{\partial z^2}$ operator in front of the integral over $\mathbb{D}$ and replace it by a finite-difference operator. The circular convolution is computed using $256\times256\times256$ FFT samples and a fixed number of 200 iterations is used in the BiCGSTAB method, which for the current QSM problem is typically sufficient to obtain accurate field or susceptibility approximations provided the BiCGSTAB-method converges, of course.

\subsection{Some simple test cases}

\begin{figure}[b]\centering
\includegraphics[scale=0.8, viewport = 10 260 700 590,clip =true]  {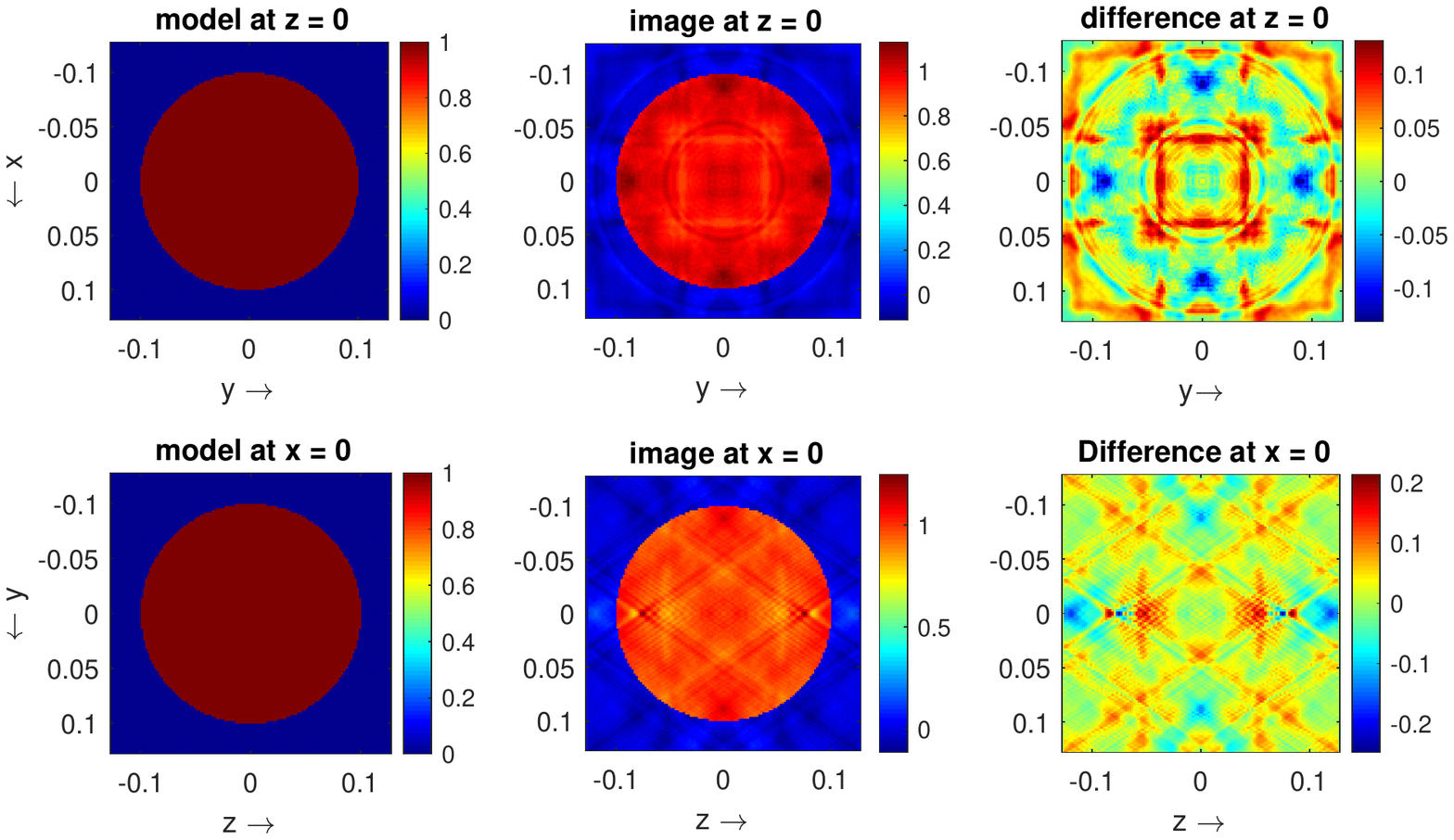}\\[-2mm]
\caption{QSM method using numerical data: Exact susceptibility~$\chi$ (left column), reconstructed susceptibility $\chi_{\rm map}$ (middle column), and  the difference $\chi\!-\!\chi_{\rm map}$ (right column). The mean error is 16\%. }
\label{Figure2_200}
\end{figure}

\begin{figure}[t]\centering
\includegraphics[scale=0.8, viewport = 10 260 700 590,clip =true]  {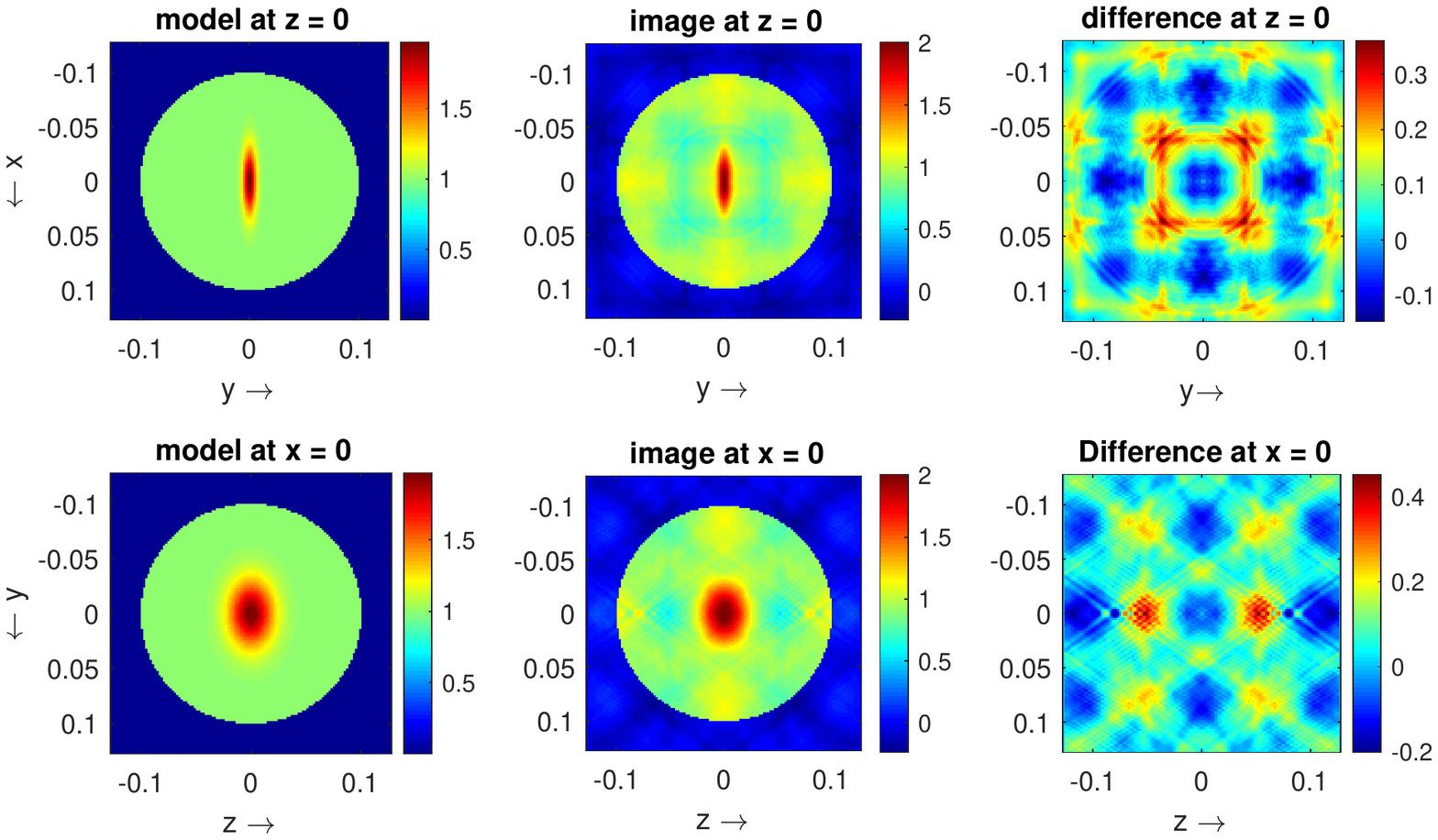}\\[-2mm]
\caption{QSM method using numerical data: Exact susceptibility~$\chi$ (left column), reconstructed susceptibility $\chi_{\rm map}$ (middle column), and difference $\chi\!-\!\chi_{\rm map}$ (right column). The mean error is 6.7\%. }
\label{Figure3_200}
\end{figure}

\begin{figure}[t]\centering
\includegraphics[scale=0.8, viewport = 10 260 700 590,clip =true]  {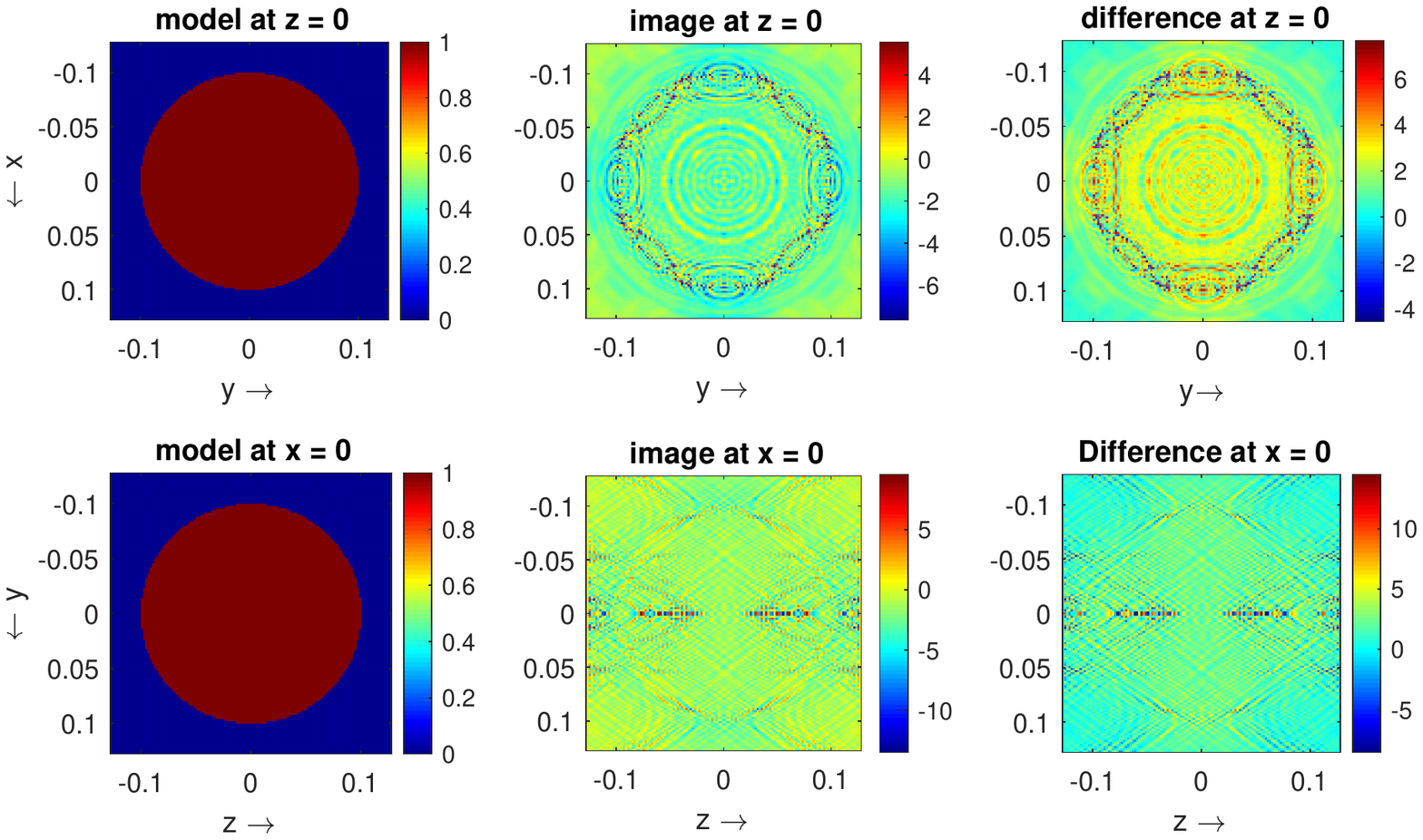}\\[-2mm]
\caption{QSM method using analytical data: Exact susceptibility~$\chi$ (left column), reconstructed susceptibility $\chi_{\rm map}$ (middle column), and difference $\chi\!-\!\chi_{\rm map}$ (right column). The mean error is 142\%.}
\label{Figure1_200}
\end{figure}

First, we consider a discretised homogeneous sphere with radius $a=10$ cm and susceptibility $\chi =1$, and compute synthetic data using (\ref{eq:QSMspacemod}). Subsequently, we attempt to retrieve the susceptibility from this data again using  (\ref{eq:QSMspacemod}).
In figure \ref{Figure2_200}, we show the exact model images $\chi$ (left column), the reconstructed images $\chi_{\rm map}$ (middle column), and their difference $\chi\!-\!\chi_{\rm map}$ (right column).
In the first row we present the susceptibilities in the cross-sectional plane $z=0$.
Here, the discrepancies between the exact and reconstructed susceptibility are around 10\%.
In the second row we present the corresponding values at the cross-section $x=0$.
These differences are enlarged by the presence of the ${\partial^2}/{\partial z^2}$ operator in the $z$-direction.
In the caption of the figure, we also present the mean error over the 3D domain $\mathbb{D}$, defined as
$
{\rm mean} \, {\rm error} = {\|\chi\!-\!\chi_{\rm map}\|_{\mathbb{D}}}\, / \,
{\|\chi\|_{\mathbb{D}}}$, where $\| \cdot \|_{\mathbb{D}}$ is the Euclidean norm on $\mathbb{D}$.
We note that increasing the number of BiCGSTAB iterations essentially does not lead to any further improvements in the reconstructions. We can conclude that even in an {\em ``inverse crime''} situation, application of the standard QSM method in the spatial domain leads to reconstruction artefacts, which are also documented in the QSM literature (see e.g. \cite{Choi}).

Second, we again consider a homogeneous sphere, but this time a defect is present inside the sphere. The defect is modeled as a smooth ellipsoid, where the susceptibility is given by
$\chi_{\rm defect}(\br) =1 + \exp(-x^2/p_x^2 -y^2/p_y^2 - z^2/p_z^2)$, with
$p_x=0.03$, $p_y=0.006$, and $p_z=0.02$.
The imaging results are presented in figure \ref{Figure3_200}.
Similar artefacts are detected, but the reconstructed defect is clearly visible, although the shape is slightly distorted.

As a final test case, we use analytical field data for the homogeneous sphere. In particular, for a sphere with radius~$a$ and a susceptibility $\chi=1$, equation (\ref{eq:QSMspacemod}) evaluates to (see (16) of \cite{Salomir}, and \ref{AppendixA})
\begin{equation}
\label{eq:QSMexample}
\Delta B(\br) =
\frac{a^3}{3}\,\frac{3z^2 -|\br|^2}{|\br|^5}, \hbox{  for} \ |\br|> a \ \quad {\rm and} \quad \Delta B(\br) = 0, \hbox{  for} \ |\br|< a.
\end{equation}
Then, for these analytical data, the iterative inversion of the susceptibility from (\ref{eq:QSMspacemod}) diverges.
The resultant image after 200 iterations is shown in figure \ref{Figure1_200}. This confirms that the inversion in the spatial domain is ill-posed as well. In this regard, we mention that the second equation of (\ref{eq:QSMexample}) shows that for a sphere, any constant susceptibility is in the nullspace of the QSM operator
\begin{equation}
\label{eq:QSMop}
A
\chi =
{\Xint-\!\!}_{\br'\in \mathbb{D}}\; d(\br-\br') \,\chi(\br')\, {\rm d}V
\end{equation}
for $\br \in \mathbb{D}$. Furthermore, Choi \etal \cite{Choi} stated that the solution of the QMS problem is a linear superposition of the actual susceptibility corresponding to the compatible portion of the data plus artefacts generated by the compatibility violators. They concluded that proper exclusion of data compatibility violations may be necessary.
An obvious culprit in this case is that $\Delta B(\br)$ in (\ref{eq:QSMexample}) jumps at the poles of the sphere ($|\br|=a$ and $z=\pm a$ in (\ref{eq:QSMexample})), which is contrary to macroscopic Maxwell theory. In absence of magnetic surface sources, the normal component of the magnetic flux density at an interface should be continuous.

In this paper we propose a different inversion strategy. Instead of mapping the susceptibility, we propose to reconstruct the magnetisation within the domain of interest from a knowledge of magnetic flux density. In the next section, we start with classical Maxwell theory and formulate this reconstruction approach.

\section{Quantitative magnetisation mapping}
 \label{sec:QMM}

In a source-free smoothly varying medium, a static magnetic field satisfies the homogeneous Maxwell equations
\begin{equation}
\label{eq:Maxwell}
\bnabla \times \bH = {\bf 0}  \quad \hbox{and} \quad  \bnabla \cdot \bB = 0,
\end{equation}
with magnetic field $\bH$ and magnetic flux density $\bB$. In case the medium parameters are not smooth and jump across a source-free interface, these equations have to be supplemented by boundary conditions. In particular, the tangential components of $\bH$ and the normal components of $\bB$ must continuous upon crossing the interface.

The relation between the magnetic flux density and the magnetic field is given by the relation
\begin{equation}
\label{eq:ConstRelation}
\bB = \mu_0 \bH + \mu_0\bM,
\end{equation}
where $\mu_0$ is the permeability of vacuum. The magnetisation is related to the magnetic field, that is, $\bM=\bM(\bH)$ and a constitutive relation describes how exactly $\bM$ depends on the magnetic field $\bH$. This relation obviously depends on the magnetic properties of the medium, but we leave this dependence unspecified for the moment.

We start by defining the background field as the field that is present when magnetisation effects are ignored. This field is denoted by $\{\bH^{\rm b}(\br),\bB^{\rm b}(\br)\}$ and for this field we have the constitutive relation $\bB^{\rm b}=\mu_0 \bH^{\rm b}$. Subsequently, we introduce the total field as the field that is present in case magnetisation effects are taken into account. At each location the total field is denoted by $\{\bH(\br),\bB(\br)\}$. In case $\bM$ does not vanish on a bounded domain $\mathbb{D}$, the difference between the total magnetic field and the background field is given by the integral equation, see e.g. \cite{Friedman},
\begin{equation}
\label{eq:Hfield}
\bH(\br)-\bH^{\rm b}(\br) = \bnabla \bnabla \cdot
\int_{\br' \in \mathbb{D} }
\frac{1}{4\pi |\br -\br'|}
 \bM(\br')
\, {\rm d}V,
\end{equation}
which holds for $\br \in \mathbb{R}^3$. If the data $[\bH(\br)-\bH^{\rm b}(\br)]$ is known for $\br \in \mathbb{D}$ (that is, data inside the domain $\mathbb{D}$ is known), Friedman \cite{Friedman} has proven that in the Hilbert space of vector functions $\mathbf{L}^2(\mathbb{D})$, a unique solution for the magnetisation $\bM$ exists.
We note that at this point the microscopic constitutive model is not important.

To check the continuity of the tangential components of the magnetic field at an interface where the magnetisation jumps, we apply the curl-operator to this integral equation. We observe that $\bnabla\times\bH=\bnabla\times\bH^{\rm b}$, since the curl of the gradient term vanishes on either side of the interface. Hence, at an interface with normal vector $\bnu$, the tangential components of the total magnetic field $\bnu\times\bH$ are continuous, since the tangential components of the background magnetic field $\bnu\times\bH^{\rm b}$ are continuous.

In QSM practice, magnetic flux densities are measured instead of magnetic field values. We therefore rewrite (\ref{eq:Hfield}) in terms of the magnetic flux density. To this end, we multiply both sides of (\ref{eq:Hfield}) by $\mu_0$ and use relation (\ref{eq:ConstRelation}) to arrive at the integral representation
\begin{equation}
\label{eq:Bfield}
\bB(\br)-\bB^{\rm b}(\br) = \mu_0\bM(\br) + \mu_0\bnabla \bnabla \cdot
\int_{\br' \in \mathbb{D} }
\frac{1}{4\pi |\br -\br'|}
 \bM(\br')
\, {\rm d}V,
\end{equation}
which holds for $\br \in \mathbb{R}^3$. For known data $[\bB-\bB^{\rm b}]$ in $\mathbb{D}$, (\ref{eq:Bfield}) is an integral equation for the unknown magnetisation  $\bM$ in $\mathbb{D}$.
Since (\ref{eq:ConstRelation}) is a linear relation and (\ref{eq:Hfield}) has a unique solution,
there exist a unique solution of (\ref{eq:Bfield}).

To check the continuity of the normal components of the magnetic flux density, we use the relations
\begin{equation}
 \bnabla\bnabla\cdot \frac{\bM(\br')}{4\pi |\br -\br'|}=\bnabla\times
\left[\bnabla\times \frac{\bM(\br')}{4\pi |\br -\br'|}
\right]
+ \bnabla^2 \frac{\bM(\br')}{4\pi |\br -\br'|}
\end{equation}
and $\bnabla^2 (4\pi |\br -\br'|)^{-1} = -\delta(\br -\br')$ with $\delta$ the Dirac function. With the help of these relations, the integral representations for the magnetic flux density can also be written as
\begin{equation}
\label{eq:Brep}
 \bB(\br)-\bB^{\rm b}(\br) =  \mu_0\bnabla\times \left[\bnabla \times
\int_{\br'\in{\mathbb{D}} }
\frac{1}{4\pi |\br -\br'|}
\bM(\br')
\,{\rm d}V\right].
\end{equation}
At an interface (with normal vector $\bnu)$ where the magnetisation jumps, continuity of the normal component of $\bB$ can now be verified by applying the div-operator to the representation of (\ref{eq:Brep}) for observation points located on either side of the interface. This gives $\bnabla\cdot\bB=\bnabla\cdot\bB^{\rm b}$, since the divergence of the curl vanishes at both sides of the interface. From this observation, we conclude that at an interface where the magnetisation $\bM$ jumps, the normal component of the known magnetic flux density $\bnu \cdot \bB$ is continuous, since the normal component of the background field $\bnu\cdot\bB^{\rm b}$ is continuous.

Finally, in QSM only the $z$-component of $\Delta \bB$ is known through measurements, while polarisation effects are neglected. In our formulation we take this into account as well, and rewrite the magnetisation as $\bM= M\bi_z$. Ignoring polarisation effects, the vector integral representations of (\ref{eq:Hfield}) and (\ref{eq:Bfield}) then simplify to the scalar equations
\begin{equation}
\label{eq:Hfieldz}
\Delta H(r)= \frac{\partial^2}{\partial z^2}
\int_{\br' \in \mathbb{D} }
\frac{1}{4\pi |\br -\br'|}
 M(\br')
\, {\rm d}V
\end{equation}
and
\begin{equation}
\label{eq:Bfieldz}
\Delta B(\br) =
M(\br) + \frac{\partial^2}{\partial z^2}
\int_{\br' \in \mathbb{D} }
\frac{1}{4\pi |\br -\br'|} M(\br')\, {\rm d}V,
\end{equation}
in which $\Delta H =
H_z(\br)-H_z^{\rm b}(\br)$ and  $\Delta B =[B_z - B_z^{\rm b}]/\mu_0$.
Note that the definition of $\Delta B$ is slightly different from the one defined in the QSM model.
In \ref{AppendixA} we show that for the special case of a homogeneous sphere, the tangential component of the magnetic field and the normal component of magnetic flux density are indeed continuous across the surface of the sphere when the above data models are used. Our objective is now to reconstruct the magnetisation $M$ from knowledge of $\Delta B$ and we refer to this inverse problem as quantitative magnetisation mapping or QMM.

\subsection{Some simple test cases}

In this section, we consider the same test cases as discussed for the QSM model, but
instead of using QSM equation (\ref{eq:QSMspacemod}), we now consider the QMM equation (\ref{eq:Bfieldz}). Again, we use the BiCGSTAB method to solve the latter equation.

\begin{figure}[t]\centering
\includegraphics[scale=0.8, viewport = 10 260 700 590,clip =true]  {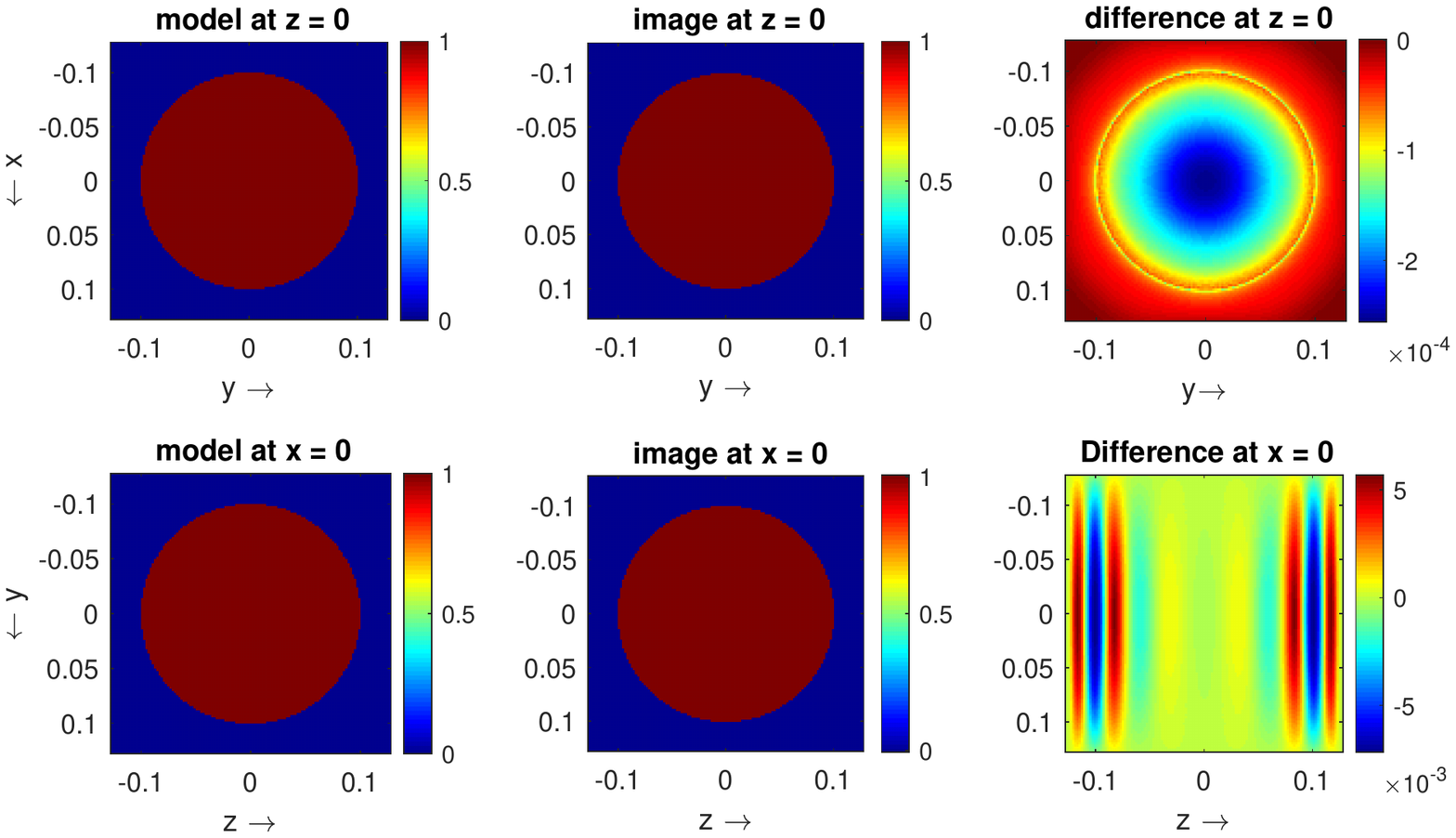}\\[-2mm]
\caption{QMM method using numerical data: Exact magnetisation~$M$ (left column), reconstructed magnetisation $M_{\rm map}$ (middle column), and  the difference $M\!-\!M_{\rm map}$ (right column). The mean error is 0.3\%. }
\label{Figure2}
\end{figure}

First, we consider the discretised homogeneous sphere  with radius $a = 10$ cm and magnetisation $M =1$, and calculate numerically synthetic data using (\ref{eq:Bfieldz}). Subsequently, we use this data and the same model (\ref{eq:Bfieldz}) to retrieve the magnetisation. In this case,
only 13 iterations are needed to reach an error criterion of $10^{-4}$. Per iteration, the computation time on a Dell Precision Tower 5810 (Single core, 3.5 GHz, 64 GB) amounts to 1.1 second.
The results are shown in figure \ref{Figure2}. In the caption of the figure, we also present the mean error over the 3D domain $\mathbb{D}$, defined as
$
{\rm mean} \, {\rm error} = {\|M\!-\!M_{\rm map}\|_{\mathbb{D}}}\, / \,
{\|M\|_{\mathbb{D}}}$.
We obtain almost perfect images with a mean error of 0.3\%.
Comparing figures (\ref{Figure2_200}) and (\ref{Figure2}), we observe that the discrepancies in the two reconstructions have a different structure. The discrepancies in QMM are also much smaller than in QSM and are not even visible on the scale of the images of the reconstructed magnetisation. Furthermore, the QMM reconstruction errors are somewhat larger in the $x$-plane than in the $z$-plane and are visible as vertical lines due to the finite difference implementation of the  ${\partial^2}/{\partial z^2}$ operator. Increasing the number of BiCGSTAB iterations from 13 to 20, the mean error is reduced further to 0.03\%.

\begin{figure}[t]\centering
\includegraphics[scale=0.8, viewport = 10 260 700 590,clip =true]  {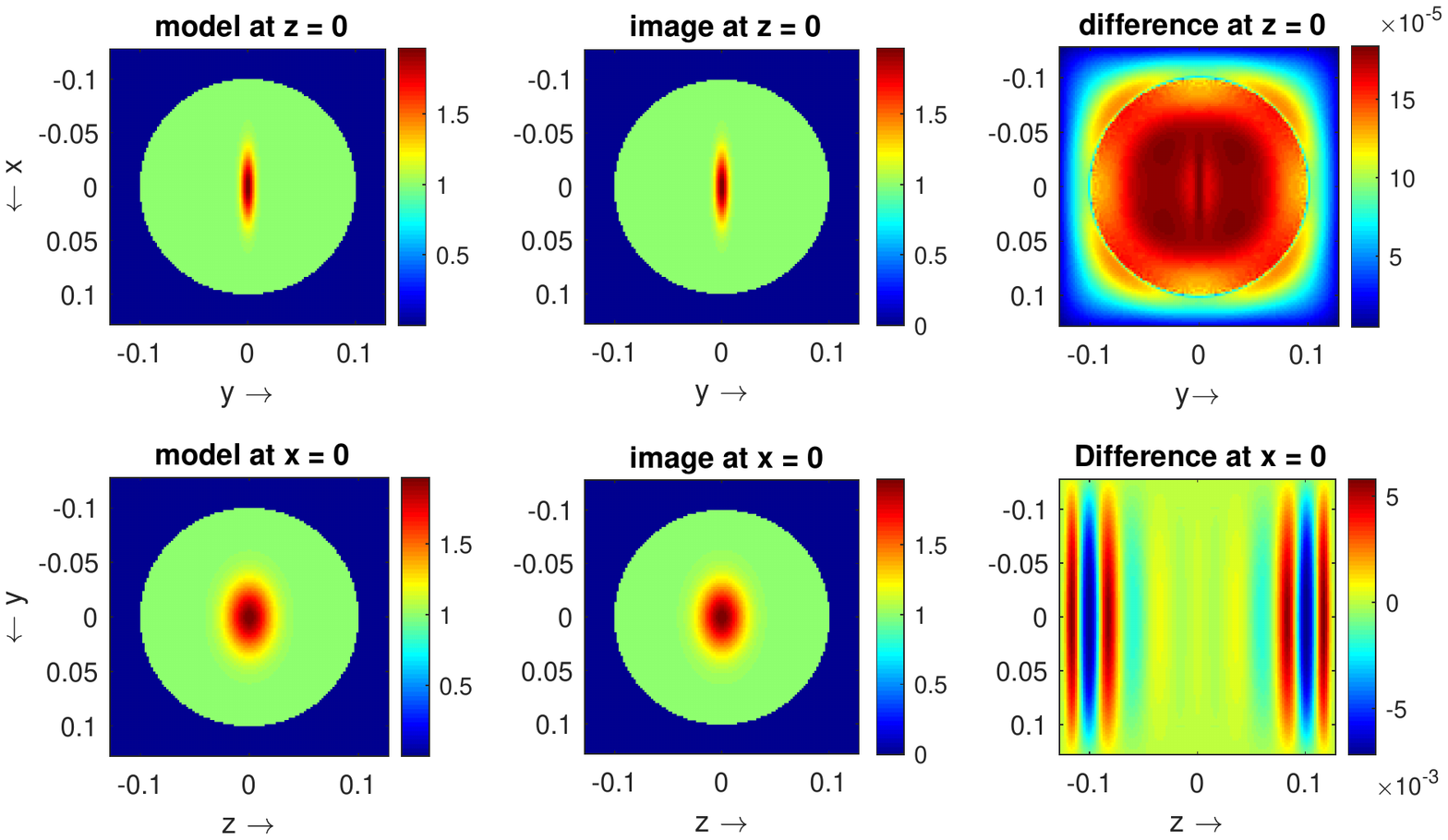}\\[-2mm]
\caption{QMM method using numerical data: Exact magnetisation~$M$ (left column), reconstructed magnetisation $M_{\rm map}$ (middle column), and difference $M\!-\!M_{\rm map}$ (right column). The mean error is 0.3\%. }
\label{Figure3}
\end{figure}

Second, we consider the homogeneous sphere with the spheroidal defect. The imaging results are presented in figure \ref{Figure3}.
Again, these results were obtained within 13 iterations and the
mean error did not change, which confirms our expectation that the resolution achieved is the same over the whole observation domain.

\begin{figure}[t]\centering
\includegraphics[scale=0.8, viewport = 10 260 700 590,clip =true]  {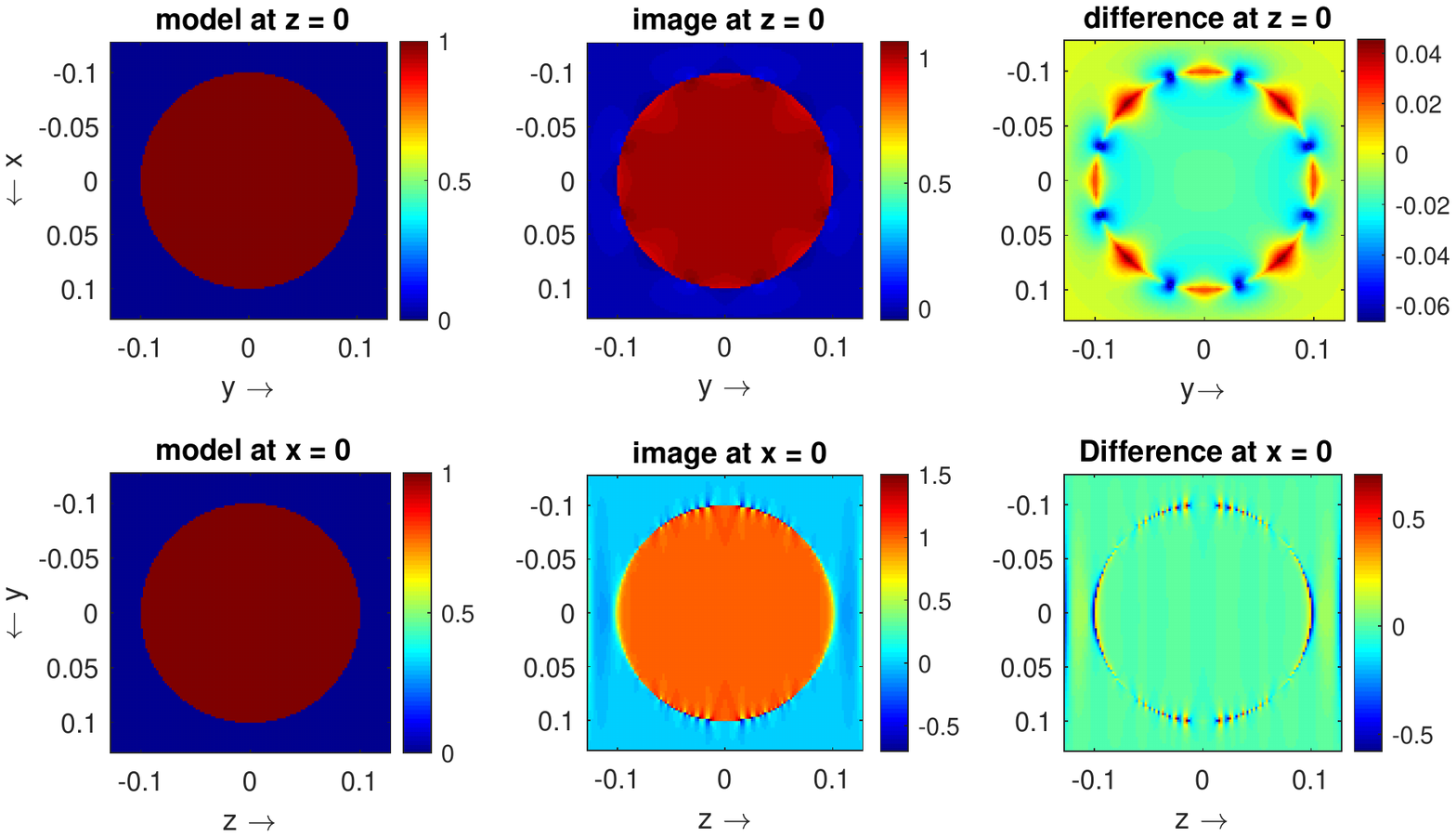}\\[-2mm]
\caption{QMM method using analytical data: Exact magnetisation~$M$ (left column), reconstructed magnetisation $M_{\rm map}$ (middle column), and difference $M\!-\!M_{\rm map}$ (right column). The mean error is 7\%.}
\label{Figure1}
\end{figure}

Third, we use the analytical data for the homogeneous sphere.
In contrast to the divergent results of figure \ref{Figure1_200}, for an error criterion of $10^{-4}$, the  BiCGSTAB method converges within 13 iterations. The results are shown in figure \ref{Figure1}.
Here, errors of half the maximum value of the model magnetisation are visible, but only at the interface of the sphere. The method shows the mismatch between the stair-case approximation and the exact spherical boundary.
These differences are magnified by the presence of the ${\partial^2} / {\partial z^2}$ operator. We further note that increasing the number of iterations reduces the residual error in the equation in question, but the mean error between $M$ and $M_{\rm map}$ does not significantly decrease.
We also remark that, in this example, the jump in the magnetisation is large compared to magnetisation changes encountered in practice. For smaller magnetisation variations, the finite difference approximation of the ${\partial^2} / {\partial z^2}$ operator is typically much better.

\begin{figure}[t]\centering
\includegraphics[scale=0.8, viewport = 10 260 700 590,clip =true]  {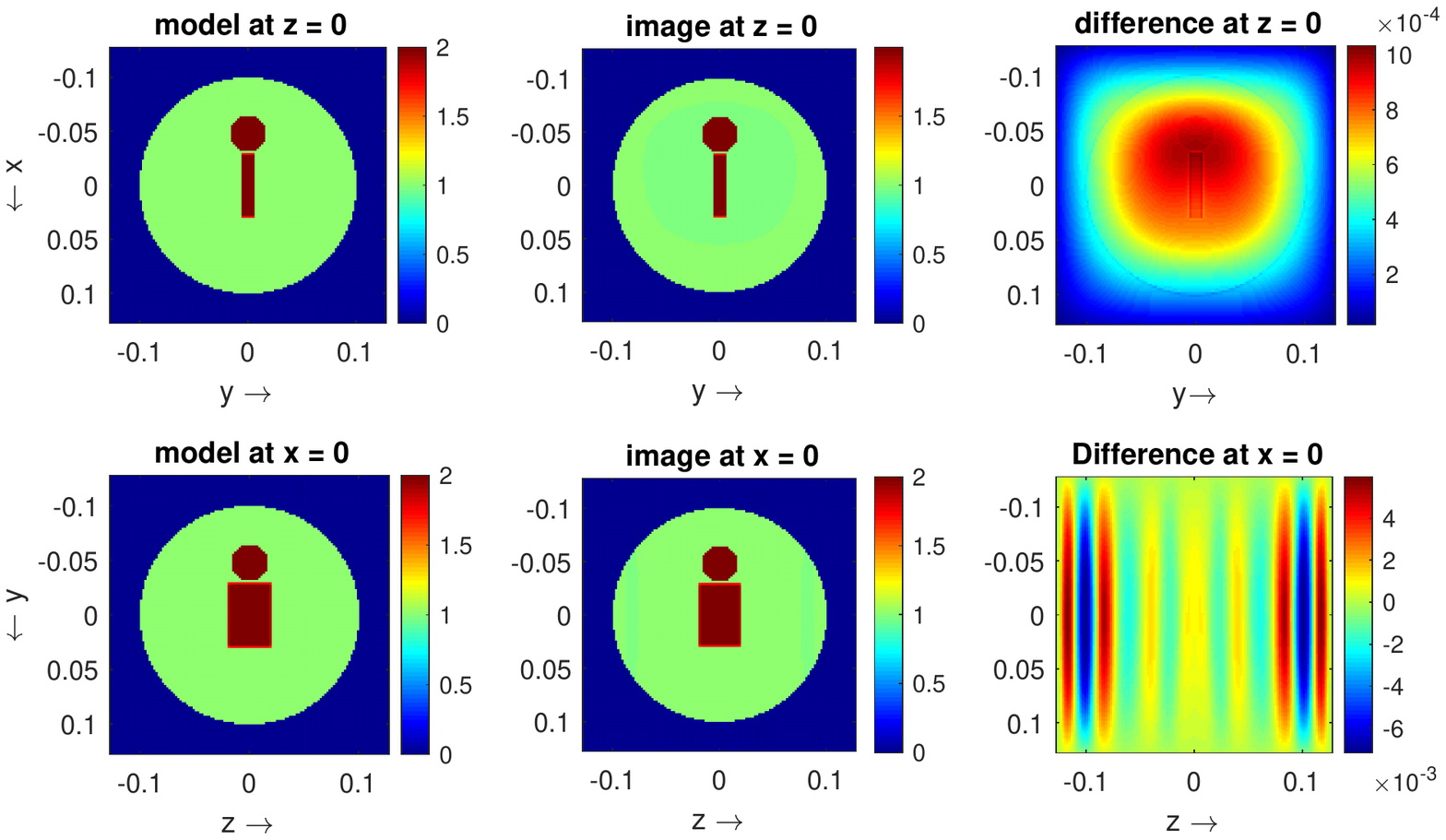}\\[-2mm]
\caption{QMM method using numerical data: Exact magnetisation~$M$ (left column), reconstructed magnetisation $M_{\rm map}$ (middle column), and difference $M\!-\!M_{\rm map}$ (right column). The mean error is 0.3\%.}
\label{Figure4}
\end{figure}

Our final example was not considered before and consists of a small sphere and a rectangular shape with smooth edges embedded in a larger sphere as illustrated in figure~\ref{Figure4}. The distance between the small sphere and the rectangular shape is equal to the sampling width of 2 mm. The reconstructed magnetisation is shown in the middle column of figure~\ref{Figure4} and the difference between the exact and reconstructed magnetisation is shown in the third column. We observe that without noise and other modelling errors, QMM is able to provide high-resolution reconstructions. Reducing the sample width results in reconstructions with an even higher resolution, although at the expense of increasing computer memory and computation time.

\section{Constitutive relation}
\label{sec:const}

As soon as the magnetisation $\bM = M\,\bi_z$ has been determined, the magnetic field $\bH$ can be calculated from (\ref{eq:Hfield}) and the constitutive parameter of the material (susceptibility) can be estimated by minimising some norm of the difference between $\bM$ and $\bH$ on the domain $\mathbb{D}$. For an isotropic medium, the constitutive relation $\bM = \chi \bH$, with $\chi$ the susceptibility, can only be satisfied when $\bH = H_z \bi_z$. Fortunately, the magnetisation in QSM practice is much smaller than the background magnetic field, and one can make the approximation $\bH \approx H_z^{\rm b}\, \bi_z$. Assuming that $H_z^{\rm b} \neq 0$, the susceptibility can then be obtained as
\begin{equation}
\chi(\br) = M(\br) / H_z^{\rm b}(\br).
\end{equation}
In principle, this is the effective susceptibility at the observation point $\br$.
When detailed information on a microscopic scale is desired, the
constitutive relation of the material or tissue is required. This relation describes the response of the material to the local magnetic field. Strictly speaking, finding such a relation requires a quantum mechanical treatment and after averaging over a suitable domain, the continuum approximation of $\bM(\br)$ at the observation point $\br$ is obtained. In other words, \emph{a priori} knowledge about tissue is required for inversion of a constitutive model. The question that remains is whether \emph{a priori} knowledge of the microscopic model is required for medical diagnostics, and if so, which model should be used for a certain material or tissue type.

\section{Comparison of the QSM and QMM operator equations}
\label{sec:compare}

In QSM, the basic imaging equation can be written as
\begin{equation}
\label{eq:Aeq}
A \varphi =f,
\end{equation}
where $\varphi$ is the unknown susceptibility $\chi$, $f$ is the data $\Delta B$ of QSM, and $A$ is the QSM operator given by (\ref{eq:QSMop}). Finding $\varphi$ from  knowledge of $f$ in $\mathbb{D}$ is an ill-posed problem and requires regularisation~\cite{Kress}. Examples of regularisation approaches are Tikhonov regularisation \cite{Tikhonov}, total variation minimisation \cite{Rudin}, or regularisation by analytical continuation \cite{Natterer}.

In terms of operator~$A$, the QMM equation of interest is given by
\begin{equation}
\label{eq:First}
\left(
\frac{2}{3}I + A\right)
\varphi = f,
\end{equation}
where this time $\varphi$ is the magnetisation $M$ and $f$ is the data $\Delta B$ of QMM. We observe that the QMM operator $\frac{2}{3} I + A$ is actually a shifted version of the QSM operator with a shift given by the scaled identity operator $\frac{2}{3}I$. The ill-posed nature of QSM can be traced back to the underlying physical model in which Lorentz correction annihilates this shift and, loosely speaking, regularisation in QSM can be seen as an attempt to restore the shift. Since such an approach may introduce regularisation artefacts and is not necessary in QMM, we propose magnetisation imaging as an alternative to QSM.

\section{Conclusions}
\label{sec:concl}

It is well known that in QSM the problem of retrieving the susceptibility from knowledge of the ($z$-component of the) magnetic flux density inside the domain of interest is an ill-posed problem. In this paper we have shown that, in addition to this ill-posedness, the magnetic flux density within the QSM model also does not satisfy the appropriate boundary condition that should hold at an interface between two media with a different magnetisation.

To avoid some of the difficulties associated with QSM, we have proposed an imaging procedure that provides maps of the magnetisation instead of the susceptibility within a region of interest. We call this imaging procedure quantitative magnetisation mapping or QMM and we have shown that within the QMM model, the boundary conditions at the interface between two different magnetic media for the magnetic field $\bH$ and magnetic flux density $\bB$ are satisfied. Moreover, a comparison of QMM and QSM shows that the QMM operator is actually a shifted version of the QSM operator. More precisely, the operator of QSM is an integral operator of the first kind, while the QMM operator is an integral operator of the second kind~\cite{Kress}. Numerical experiments for elementary three-dimensional structures have also been presented and show that in QMM the magnetisation can be retrieved in just a few iterations of an iterative solver such as BiCGSTAB. Furthermore, our simple imaging examples exhibit very high resolution over the whole window of observation, which may have significant consequences for medical diagnostics.

With a magnetisation map $M$ at our disposal, the corresponding magnetic field $H$ can be determined from the integral representation that relates the magnetic field to the magnetisation. Subsequently, if a parametric model $M(H)$ is known from either microscopic or quantum mechanical considerations, then the model parameters can be determined by matching the model to the reconstructed magnetisation. Obviously, the mapping procedure in QMM does not depend on a particular material or tissue model, since it directly reconstructs the magnetisation from measured field data.

Finally, we mention that in this paper we have focused on the basic QSM and QMM equations and we used simple 3D structures. We used either simulated or analytic data sets to demonstrate some of the fundamental properties of the QSM and QMM operators. In practice, however, we have to deal with measured data sets containing noise and other (background) field perturbations. The performance of QMM on such data sets will be investigated in future work.

\clearpage

\appendix

\section{Homogeneous sphere in a uniform field}
\label{AppendixA}
To check the continuity of the tangential magnetic field and the normal magnetic flux density,
we consider the special case of a homogeneous sphere that occupies the spherical domain $\mathbb{B}$ with radius $a$ and center at the origin $\br=\bf 0$.
We assume that, inside the sphere, the susceptibility $\chi$ is constant in the QSM model, and the magnetisation $M$ is constant in the QMM model. For both  models, we use the relation
\begin{equation}
\label{eq:dzdzint}
{\cal K}(\br)=\frac{\partial^2}{\partial z^2}
 \int_{\br' \in \mathbb{B} }
\frac{1}{4\pi |\br-\br'|}\, {\rm d}V = \left\{
  \begin{array}{ll}
   \displaystyle
 \frac{a^3}{3}\,\frac{3z^2 -|\br|^2}{|\br|^5},   &\hbox{for } |\br| > a,\\[2mm]
       \displaystyle                      - \frac{1}{3}, &\hbox{for } |\br| <  a.                                            \end{array}
                                          \right.
\end{equation}
At the poles of the sphere, $|\br| = a$ and $z =\pm \, a$ in (\ref{eq:dzdzint}),  and ${\cal K}$ is discontinuous in the radial direction. When we approach the poles, the limiting values of ${\cal K}$ are
\begin{equation}
\lim_{|\br| \,\downarrow\, a}{\cal K}(\br) = \frac{2}{3} \quad {\rm and} \quad
\lim_{|\br|\, \uparrow \, a }{\cal K}(\br)      = -\frac{1}{3}.
\end{equation}
At the equator of the sphere, $|\br| = a$ and $z=0$ in (\ref{eq:dzdzint}),  and ${\cal K}$ is continuous in the radial direction, where ${\cal K} = -1/3$.

\vspace{5mm}
\noindent
For the {\em QSM model}, we start with (\ref{eq:QSMspacemod}). Here, the magnetic flux density is obtained as
\begin{equation}
\Delta B = {\cal K(\br)}\,\chi, \quad \hbox{for $|\br| > a$},   \quad  {\rm and} \quad
\Delta B = 0, \ \hbox{ for $|\br| <  a$}.
\end{equation}
At the poles of the sphere, the limiting values of the magnetic flux density are
\begin{equation}
\lim_{ |\br|\, \downarrow\, a}\Delta B(\br) = \frac{2}{3}\chi  \quad {\rm and} \quad
\lim_{|\br|\, \uparrow  \,  a}\Delta B(\br) = 0.
\end{equation}
At these pole locations, $\Delta B$ is pointing in a direction normal to the sphere, but it is not continuous. The jumps amounts to $2\chi/3$. This contradicts the macroscopic theory of Maxwell's equations.

\vspace*{5mm}
\noindent
For the {\em QMM model}, we start with (\ref{eq:Hfieldz}) and (\ref{eq:Bfieldz}). The magnetic field is obtained as
\begin{equation}
\label{eq:homsphereH}
\hspace{-1cm}
\Delta H(\br) = {\cal K(\br)}\,M, \ \hbox{for $|\br| > a$},   \ \ {\rm and} \ \
\Delta H(\br) =  {\cal K(\br)}\,M, \ \hbox{for $|\br| <  a$}.
\end{equation}
 At the equator of the sphere, ${\cal K}(\br)$  is continuous in radial direction and we observe that the tangential component of the magnetic field is indeed continuous.
Furthermore the magnetic flux density is given by
\begin{equation}
\hspace{-1cm}
\label{eq:homsphereB}
\Delta B(\br) = {\cal K(\br)}\, M, \ \hbox{for $|\br| > a$},  \ \ {\rm and}  \  \
\Delta B(\br) = M+{\cal K(\br)}\, M, \ \hbox{for $|\br| <  a$}.
\end{equation}
At the poles of the sphere, the limiting values of the magnetic flux density are
\begin{equation}
\lim_{|\br|\, \downarrow \, a}\Delta B(\br) = \frac{2}{3} M  \quad  {\rm and} \
\lim_{|\br|\, \uparrow \, a}\Delta B(\br) = \frac{2}{3} M,
\end{equation}
which shows that the normal component of the magnetic flux density is continuous as well.
Obviously, the QMM model is compatible with the macroscopic Maxwell equations.

%
% REFERENCES
%

\end{document}